# Effect of Short-Term Debt on Financial Growth of Non-Financial Firms Listed at Nairobi Securities Exchange

David Haritone Shikumo[1]    Dr. Oluoch Oluoch[2]    Dr. Joshua Matanda Wepukhulu[3]
1. PhD Finance Candidate
Jomo Kenyatta University of Agriculture and Technology, Kenya
2. Senior Lecturer - Department of Commerce and Economic Studies
Jomo Kenyatta University of Agriculture and Technology, Kenya
3. Lecturer - Department of Commerce and Economic Studies
Jomo Kenyatta University of Agriculture and Technology, Kenya

**Abstract**
A significant number of the non-financial firms listed at Nairobi Securities Exchange (NSE) have been experiencing declining financial performance which deter investors from investing in such firms. The lenders are also not willing to lend to such firms. As such, the firms struggle to raise funds for their operations. Prudent financing decisions can lead to financial growth of the firm. The purpose of this study is to assess the effect of short-term debt on financial growth of non-financial firms listed at Nairobi Securities Exchange for a period of ten years from 2008 to 2017. Financial firms were excluded because of their specific sector characteristics and stringent regulatory framework. The study is guided by Agency Theory and Theory of Growth of the Firm. Explanatory research design was adopted. The target population of the study comprised of 45 non-financial firms listed at the NSE for a period of ten years from 2008 to 2017. The study conducted both descriptive statistics analysis and panel data analysis. The result indicates that, short term debt explains 45.99% and 25.6% of variations in financial growth as measured by growth in earnings per share and growth in market capitalization respectively. Short term debt positively and significantly influences financial growth measured using both growth in earnings per share and growth in market capitalization. The study recommends that, the management of non-financial firms listed at Nairobi Securities Exchange to employ financing means that can improve the earnings per share, market capitalization and enhance the value of the firm for the benefit of its stakeholders.
**Keywords**: Short Term Debt, Non-financial Firms, Nairobi Securities Exchange, Growth in Earnings per Share, Growth in Market Capitalization.
**DOI:** 10.7176/RJFA/11-17-16
**Publication date:** October 31st 2020

## 1.0 INTRODUCTION

Short term debt is part of the financial structure. Financial structure is the way a firm finances its assets through some combination of debt and equity that a firm deems as appropriate to enhance its operations (Kumah, 2013). The determination of a firm's optimal financial structure is vital in deciding how much money should be borrowed and the best mixture of debt and equity to fund business operations (Shubita & Alsawalhah, 2012). Therefore, the choice among ideal proportion of debt and equity can affect the value of the firm, as well as financial performance. Short-term assets and liabilities are generally defined to be those items that will be used, liquidated, mature or paid off within one year. Short-term assets should be financed with short-term liabilities (Guin, 2011). Short-term term is primarily concerned with the analysis of decisions that affect current assets and current liabilities. Short term debt is measured as short-term liabilities divided by total assets. According to Garcia-Terul and Martinez -Solano, 2008), short-term debt is positively correlated with firm's growth opportunities. The anecdotal evidence suggests that there is a positive relationship between short term debt financing and financial performance (Yazdanfar & Öhman, 2015).

Short term debt financing has a maturity period of one year or less, they must be re-paid quickly within 90 – 120 days. Term loans with short maturities help to meet immediate need for financing without long term commitment (Peavler, 2014). The cost of servicing short term debt is less taxing on the firm. Short term loans usually offer lower interest charges, and most lenders do not charge interest until all credit allowance period is breached (Kahl, Shivdasani & Wang, 2015).

Short-term assets and liabilities management are critical because they lay bare the financial stability of the firm and the market (Peirson, Brown, Easton & Howard, 2014). Short term debt has been observed by various scholars and researchers to have influence on profitability. Baum, Schaafer and Talavera (2006) found that firms can make use of short-term financing which may influence profitability of the firm depending on the cost of the source of financing to that particular firm. Baum *et al.* (2006) observed that firms may have a certain ration of short-term liabilities in its financing structure which they feel are optimum in enhancing performance and profitability. In the case of Germany, Diamond and He (2014) observed that firms which had high short-term debt levels when compared to their long-term debt performed better than their peers.

Githaiga and Kabiru (2015) measured short term debt as ratio of short-term loan to total loan. Magoro and Abeywardhana (2017) measured short term debt as the ratio of debt repayable within one year to total assets. In addition, Ma'aji, Abdullah and Khaw (2018) measured short term debt as the ratio of short-term liabilities to total assets. In this study, short term debt was measured as the ratio of current liabilities to total assets.

According to Tailab (2014) the use of short-term liabilities such as trade payables and accruals can have a positive effect on a firm's profitability since such sources of financing may be less costly to the business than the long-term sources of funds. Further, short term sources of funds may have a positive influence on profitability due to the reduced contractual engagements that are involved (Krishnamurthy & Vissing-Jorgensen, 2013). Short term credit may have a positive influence on profitability. However, García-Teruel and Martínez-Solano (2007) refutes by saying that short maturity of short-term debt may prove expensive to the firm hence increasing its cost of capital. Financial growth is a measure of efficient utilization of assets by a firm from principal business mode to generate revenues (Aburub, 2012). Financial growth is a general measure of the overall financial health of a firm over a given period (Onaolapo & Kajola, 2010). According to Buvanendra *et al.* (2017), the financial growth of a company is measured as the growth of market capitalization. Market capitalization refers to the total dollar market value of a company's outstanding shares (San & Heng, 2011). Market capitalization is calculated by multiplying a company's shares outstanding by the current market price of one share (Buvanendra, Sridharan & Thiyagarajan, 2017). The return on investment, return on assets, market value, and accounting profitability reflect financial growth of firms (Ongore & Kusa, 2013).

## 1.1 The Nairobi Securities Exchange (NSE)
The Nairobi Securities Exchange (NSE) was founded in 1954. NSE play a vital role in the growth of Kenya's economy by encouraging savings and investment, as well as helping local and international firms access cost-effective capital. NSE operates under the jurisdiction of the Capital Markets Authority of Kenya. Nairobi Securities Exchange lists both financial and non-financial firms. The non-financial firms are those firms that are not involved in the provision of financial intermediary services (NSE, 2016). The non-financial firms listed at NSE include agriculture allied firms, automobiles and accessories, commercial and services, construction and allied, energy and petroleum, manufacturing and allied, investment, telecommunication and technology and real estate investment trust (NSE, 2016). The non-financial firms listed at NSE formed the focus of the study.

## 1.2 Statement of the Problem
A significant number of the non-financial firms listed at Nairobi Securities Exchange (NSE) have been experiencing declining financial performance which deter investors from investing in such firms (Muchiri, Muturi & Ngumi, 2016). The growth of non-financial firms listed at Nairobi Securities Exchange was 3.7% in 2017 against 4.2% in 2016 (NSE, 2017). Decline in financial performance deter lenders from lending to such firms (Muchiri, Muturi & Ngumi, 2016). For instance, Kenya Airways Limited reported a net loss of Kshs. 26.2bn ($258m) for the financial year 2015-2016, up from Kshs. 25.7bn in the previous financial year (NSE, 2016). Uchumi Supermarket Limited was revived after an agreement between the Kenyan government, suppliers and debenture holders (NSE, 2017).

Majority of the studies conducted; Akbarpour and Aghabeygzadeh (2011), Muchiri, Muturi and Ngumi (2016), Shubita and Alsawalhah (2012), Habib, Khan and Wazir (2016) and Chen (2014) focused on capital structure while basing their argument on accounting concept. Unlike financial structure, short term liabilities do not contribute to capital structure (Opungu, 2016; Muchiri, Muturi & Ngumi, 2016). There are also inconsistencies of results from previous empirical studies on the effect of short-term debt on financial performance of listed non-financial firms (Menike & Prabath, 2014; Muchiri, Muturi & Ngumi, 2016). The study intends to fill this conceptual gap by focusing on the effect of short-term debt on financial growth of non-financial firms listed at Nairobi Securities Exchange.

## 1.3 Research Objective
To assess the effect of Short-term debt on financial growth of Non-financial firms listed at Nairobi Securities Exchange.

## 1.4 Statistical Hypothesis
There is no significant effect of Short-term debt on financial growth of Non-financial firms listed at Nairobi Securities Exchange.

## 2.0 LITERATURE REVIEW
### 2.1 Theoretical Review
A theory is a generalization about a phenomenon, an explanation of how or why something occurs. Theories describe, explain, predict, or control human phenomena in a variety of contexts. According to McMillan and Schumacher (2006), a theory is an explanation, a systematic account of relationships among phenomena. The study is guided by Agency Theory and Theory of Growth of the Firm.

### 2.1.1 Agency Theory

Jensen and Meckling (1976) advanced the agency theory which states that, the agency costs that arise from the conflicts between the managers and the owners of the firm are reduced by having a certain proportion of debt in the financial structure of the firm (Leland, 1998). The lowering of agency conflicts would lead to reduction in agency costs which would lead to improved financial growth. The use of debt in the firm as observed by Jensen and Meckling (1976) can help to control and monitor managers in the firm to ensure that they follow objectives that are beneficial to the firm.

Buferna, Bangassa and Hodgkinson (2005) supported this theory by indicating that inclusion of debt in the financial structure provides a motivation for managers to stimulate the growth of a firm so as to have cash flows that would satisfy repayments of debts. This leads to the enhancement of the firm's profitability (Dawar, 2014). This theory postulates that short term debt and any other form of debt that a firm uses reduces agency conflicts between managers and shareholders of the firm and hence boosts financial growth (Rashid, 2015). The agency theory plays a crucial role in financing decisions because of the problems that arise between the debt holders and the shareholders.

### 2.1.2 Theory of Growth of the Firm

The theory was propagated by Penrose (1959). Penrose argued that firms had no determinant to long run or optimum size, but only a constraint on current period growth rates (Penrose, 1959). According to the theory, financial means for expansion could be found through retained earnings, borrowing, and new issues of stock shares. Retained earnings are one of the most important sources to finance new projects in emerging economies where capital markets are not well developed. However, firms in the start-up period, when initial investments have not matured yet or whose investment projects are substantially larger than their current earnings, will not have enough financial means from retained earnings and will face a constraint in their growth project. Firms in this situation may seek external sources of financing; however, the extent of borrowing could be limited by internal factors like high debt-equity ratios that would expose both borrower and lender to increased risk. In other cases, financing of growth projects may be limited by shallow financial markets. Rajan and Zingales (1998) found that industrial sectors with a great need for external finance grow substantially less in countries without well-developed financial markets.

This theory is relevant to this study since it informs the dependent variable which is financial growth. The current studies which have used this theory of firm's growth are; Diaz Hermelo (2007) who conducted a study on the determinants of firm's growth: an empirical examination and Pervan and Višić (2012) who conducted a study on the Influence of firm size on its business success.

### 2.2 Empirical Review

Baum, Schafer and Talavera (2006) conducted a study on the effects of short-term liabilities on profitability: a comparison of German and US firms. The specific objective was to establish the effects of short-term liabilities on profitability by comparing German and US firms. The paper adopted the methodology of the empirical finance literature to analyze a common question that liability maturity structure has an impact on firm performance. A comparison is made between two countries, the US and Germany, with different types of financial systems. They study found that, German firms rely more heavily on short-term liabilities that are likely to be more profitable. The link between liability maturity structure and profitability does not appear in the results from the US sample, which reflects the importance of institutional factors.

García, Teruel, Martínez and Solano (2008) analyzed the determinants of cash holding in Spanish SMEs and found that, firms with a higher amount of short-term debt will hold higher levels of cash, because it might lower the risks of the non-renewal of the short-term debt. The specific objectives were to establish the effect short-term debt, long-term capital and tangibility on firm's profitability. The study adopted descriptive survey research design. The study established that, short-term debt is positively related to firm's profitability, which might be the most important factor in accessing outside financing in countries with weak collateral laws.

Fosberg (2013) conducted a study on short-term debt financing during the financial crisis. The financial crisis of the late 2000s had a large effect on the capital and lending markets in the United States and overseas. The data presented showed that, the financial crisis caused firms to increase the amount of short-term debt they employed from 1.3% of assets in 2006 to 2.2% in 2008. This increase in short-term debt financing was completely reversed by the end of 2009 suggesting that the increase in short-term debt financing was undesired and was reversed as soon as the financial crisis abated. The proximate causes of the spike in short-term debt financing include a reduction in accounts payable financing from suppliers and a decline in long-term debt and equity financing. A significant decrease in sale of assets also contributed to the need for more short-term debt financing. A regression analysis indicated that, the increase in short-term debt financing was caused by the financial crisis and not the simultaneous recession.

Pouraghajan, Malekian, Emamgholipour, Lotfollahpour and Bagheri (2012) conducted a study on the relationship between capital structure and firm performance evaluation measures: Evidence from the Tehran Stock Exchange. Descriptive research design was applied. A panel data model was used to data collected. The results showed that short term debt and long-term debt asset ratio do not have a significant effect on firms' profitability.

Habib, Khan and Wazir (2016) conducted a study on the impact of debt on profitability of firms; evidence from non-financial sector of Pakistan. The entire non-financial sector of Pakistan was selected for the study, but some companies were eliminated due to unavailability of data. After eliminating such companies, the data consist of 340 firms listed at Karachi Stock Exchange (KSE) for the period 2003–2012 was used for the analysis. Panel research design was employed. Return on Assets is used as a measure of profitability and is the dependent variable, whereas; short term debt to assets, long term debt to assets, total debt to assets are used as independent variables, while firm size, sales growth, and growth opportunity are used as control variables. Random effect regression analysis is used to find out the impact of debt on profitability. The results indicated a significant but negative relationship between short term debt, long term debt, total debt, and return on assets.

## 2.3 Conceptual Framework
Figure 1 shows the conceptual framework.

**Independent Variable**            **Dependent Variable**

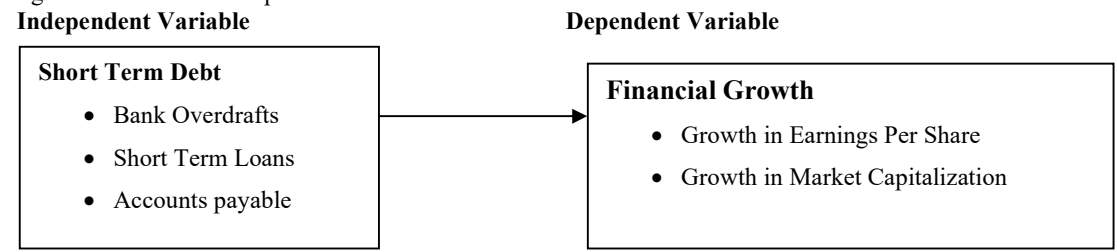

**Figure 1: Conceptual Framework of Short-Term Debt and Financial growth**

The Short-term debt is the independent variable. The dependent variable is financial growth measured using both growth in earnings per share and growth in market capitalization.

## 3.0 METHODOLOGY
The study adopted positivism research philosophy. Positivism philosophy premises that, knowledge is based on facts and that no abstractions or subjective status of individuals is considered (Bryman & Bell, 2011). Explanatory research design was also adopted. Explanatory study sets out to explain and account for the descriptive information by seek to answer the 'why' and 'how' questions (Subedi, 2016). It builds on exploratory and descriptive research and goes on to identify actual reasons a phenomenon occurs. The target population of the study comprised of 45 non-financial firms listed at the NSE for a period of ten years from 2008 to 31st December 2017 (NSE, 2017). The study adopted a census technique where all the 45 non-financial firms listed at NSE were included in the study.

Secondary data for short term debt, earnings per share and market capitalization were extracted from firms published audited financial statements and NSE handbooks. Short term debt was measured as ratio of current liabilities to total Assets, Earnings per share (EPS) was calculated as a firm's profit divided by the outstanding shares of its common stock. Market capitalization was measured as market price per share or stock multiplied with the total number of ordinary shares.

The study conducted both descriptive statistics analysis and panel data analysis. Panel analysis permits the researcher to study the dynamics of change with short time series. The combination of time series with cross-section can enhance the quality and quantity of data in ways that would be impossible using only one of these two dimensions (Gujarati, 2009). The target population of the study comprised 45 non-financial firms listed at the NSE for a period of ten years from 2008 to 31st December 2017 (NSE, 2017). Secondary data was extracted from published audited financial statements. Panel data obtained covered a period of 10 years beginning from 2008 and ending in 2017. The panel model to be estimated was: -

$FG_{it} = \beta_0 + \beta_1 STD_{it} + \mu$

Where;

FG = Financial growth measured by growth in earnings per share and growth in market capitalization of firm i at time t

$\beta_0$ = Alpha coefficient representing the constant term

$\beta_i$ = Beta coefficient

STD = Short term debt of firm i at time t

i = Firms listed from 2008 to 2017

t = Time period (2008-2017)

μ = Error term

## 4.0 RESEARCH FINDINGS AND DISCUSSIONS
### 4.1 Descriptive Statistics
Table 1 shows the descriptive statistics for short term debt, earnings per share and market capitalization.

**Table 1: Descriptive Statistics**

| Variable | Obs | Mean | Std. Dev. | Min | Max |
|---|---|---|---|---|---|
| Short Term Debt | 360 | 0.29146 | 0.255896 | 0.007901 | 2.535623 |
| EPS | 360 | 6.468265 | 15.03232 | -46.744 | 100.0483 |
| Market Capitalization in million KES | 360 | 24600.00 | 77300.00 | 116.000 | 721000.00 |

The descriptive statistic results show that, the mean value for short term debt was 0.29146 with a minimum of 0.007901 and a maximum of 2.535623. The variation in standard deviation was 0.255896. The mean value for earnings per share was 6.468265 with a minimum of -46.744, a maximum of 100.0483 and standard deviation of 15.03232. The mean value for market capitalization as another measure of financial growth was Kshs. 24600 million with a minimum of Kshs. 116 million and a maximum of Kshs. 721000 million.

### 4.2 Trend Analysis
This section presents the analysis of the trends of the variables. The study conducted a trend analysis to establish the movement of the variables overtime. Trend analysis for short term debt, earnings per share and market capitalization are presented in figure 2, figure 3 and figure 4 respectively.

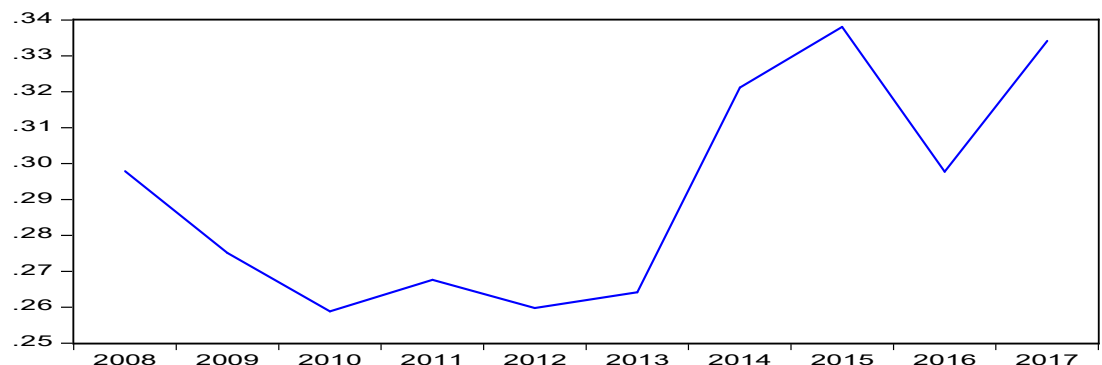

**Figure 2: Short Term Debt Trend Line**

Short term debt was lowest in 2010 before rising to reach highest value in 2015. This could imply that short-term debt financing was easily available compared to the long-term debt which is usually associated with high value collateral and at times restrictive covenants to make it unattractive. The huge proportion of asset financing through short term debt could imply that short-term debt financing was less costly and therefore available compared to the long-term debt which is usually associated with high value collateral and at times restrictive covenants making it unattractive. Short term debt is made up of any debt incurred by a firm that is due within the current fiscal year. The value of short-term debt is very important when determining a firm's financial performance. According to Muchugia (2013) there was significant positive relationship between short term debt financing and profitability because short-term debt tends to be less expensive and increasing it with a relatively low interest rate will lead to an increase in profit levels and therefore performance. These findings contradict Mwangi, Muathe and Kosimbei (2014) who concluded that majority of firms at the NSE use long term debt to finance their assets.

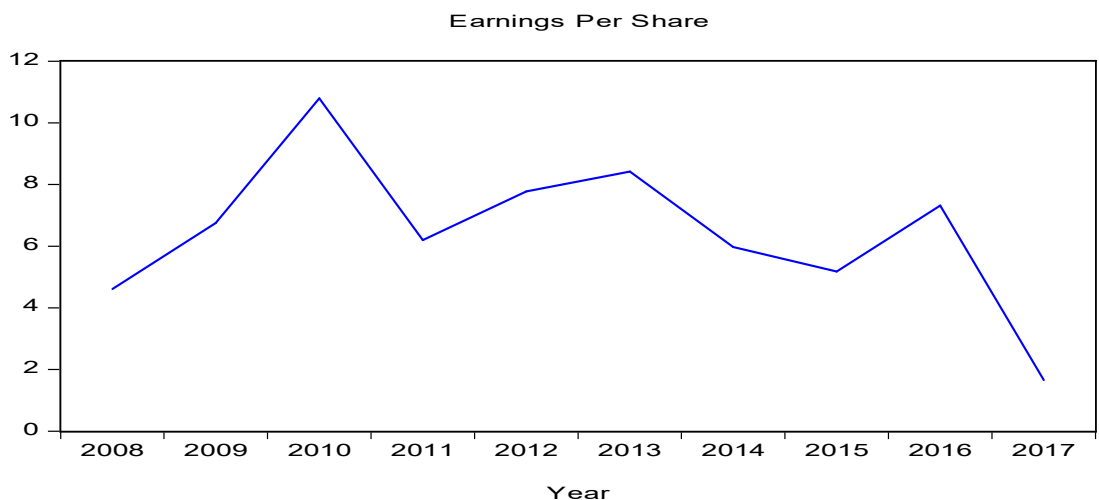

**Figure 3: Earnings per Share Trend Line**

The trend line shows that, earnings per share rose steadily from 2008 to reach highest in 2010. However, earning per share dropped further from 2011 to 2012 before rising again in 2013. Earnings per share further dropped drastically to reach lowest in 2017. Earnings per share is considered as an important accounting indicator of risk, entity performance and corporate success. It is used to forecast potential growth in future share prices, because changes in earning per share are often reflected in share price behavior. Smart and Graham (2012) concur by suggesting that an entity's growth rate is determined by performance indicators such as EPS which is disclosed in the financial statements of companies according to the specifications of the specific accounting standards applied in the respective country. Furthermore, authors have argued that, earning per share has become a useful investment decision tool for investors, because it indicates future prospects and growth (Mlonzi, Kruger & Ntoesane, 2011). According to Robbetze, de Villiers and Harmse (2017) EPS correlated best with the changing behavior of share prices.

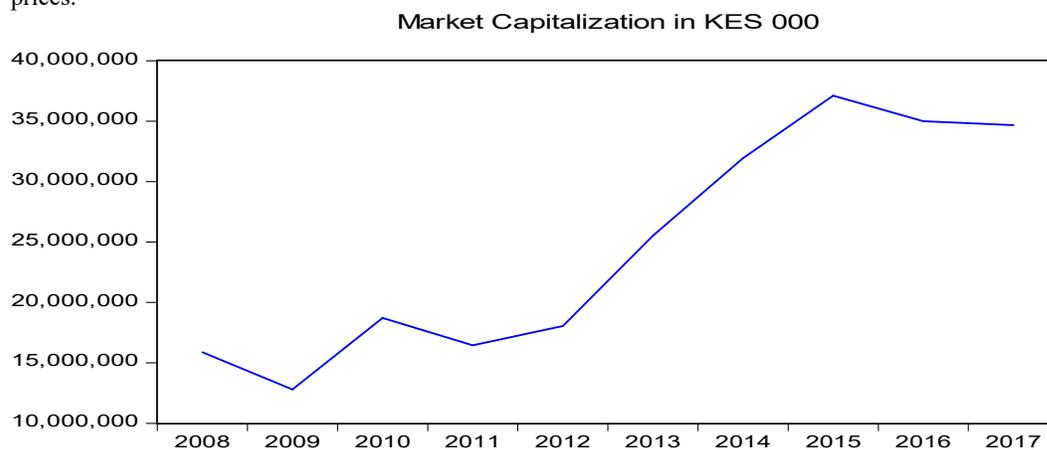

**Figure 4: Market Capitalization Trend Line**

Market capitalization was lowest in 2009 and sharply rose to highest in 2015. Market capitalization is important in projecting the size of an organization because it shows the organization's value. Market capitalization is a measure of the value of companies and stock markets which is an on-going market valuation of a public firm whose shares are publicly traded on a stock exchange computed by multiplying the number of outstanding shares held by the shareholders with the current per share market price at a given time. A market capitalization calculation is a critical part of any stock valuation formula as it represents the total market value of all the firm's outstanding shares. This represents the value the market has placed on the value of a firm's equity. As outstanding stock is bought and sold in public markets, market capitalization could be used as a proxy for the public opinion of a firm's net worth and is a determining factor in some forms of stock valuation. Market capitalization represents the public consensus on the value of a company's equity. According to Koila, Kiru and Koima (2018) using random effects model also revealed that market capitalization cannot be used to predict the outcome of return on equity within the listed firms at Nairobi Securities Exchange.

**4.3 Correlation Analysis**
Table 2 shows the correlation matrix of short-term debt and growth in earnings per share.
**Table 2: Correlation between Short Term Debt and Growth in Earnings per Share (EPS)**

|  | **Growth in EPS** | **Short Term Debt** |
|---|---|---|
| Growth in EPS | 1.000 |  |
| Short Term Debt | 0.211 | 1.000 |

The correlation results found that short term debt and growth in earnings per share are positively associated. Short-term assets and liabilities are generally defined to be those items that will be used, liquidated, mature or paid off within one year. Short-term assets should be financed with short-term liabilities and long-term assets should be financed with long-term liabilities (Guin, 2011). Short-term term is primarily concerned with the analysis of decisions that affect current assets and current liabilities. The results agree with Baum, Schafer and Talavera (2006) who conducted a study on the effects of short-term liabilities on profitability, a comparison of German and United States (US) firms and found firms rely more heavily on short-term liabilities are likely to be more profitable. Table 3 shows the correlation matrix of short-term debt and growth of market capitalization as a measure of financial growth.

**Table 3: Correlation between Short Term Debt and Growth in Market Capitalization**

|  | **Growth in Market Capitalization** | **Short Term Debt** |
|---|---|---|
| Growth in Market Capitalization | 1.000 |  |
| Short Term Debt | 0.221 | 1.000 |

The correlation results found that short term debt and growth of market capitalization are positively associated. Short-term debt could be used as permanent source of financing if the debt is continually refinanced as it matures. One reason to use short-term debt as a permanent source of financing is to take advantage of an upward sloping yield curve to reduce the firm's interest expense. The results agree with Fosberg (2013 that the amount of firm's short-term debt financing is shown to be inversely related to the amount of the firm's non-debt tax shields, growth opportunities.

**4.4 Diagnostic Tests**
**4.4.1 Stationarity Test**
Since panel data has both cross-sections and time series dimensions, there is need to test for stationarity of the time series because the estimation of the times series assumes that the variables are stationary. Estimating models without considering the non-stationary nature of the data would lead to unauthentic results (Gujarati, 2009). The study employed Fisher-type test in testing the stationarity of the data. Stationarity results are presented in Table 4. The hypotheses to be tested were;
Ho: All panels contain unit roots
Ha: At least one panel is stationary

**Table 4: Fisher-type Test of Unit Root**

| Variable |  | Inverse chi-squared (70) P | Inverse normal Z | Inverse logit t (179) L* | Modified inv. chi-squared Pm |
|---|---|---|---|---|---|
| Short Term Debt | test statistic | 99.4495 | -1.8098 | -1.8875 | 2.2875 |
|  | p-value | 0.0178 | 0.0352 | 0.0303 | 0.0111 |
| Earning Per Share | test statistic | 221.4198 | -8.991 | -9.5401 | 12.4516 |
|  | p-value | 0.000 | 0.000 | 0.000 | 0.000 |
| Market Capitalization | test statistic | 324.5335 | -12.1493 | -14.545 | 21.0445 |
|  | p-value | 0.000 | 0.000 | 0.000 | 0.000 |

The stationarity results test for unit root revealed that, at level short-term debt, earning per share and Market capitalization were stationary since p-value<0.05 at P, Z, L* and Pm. Long term debt and retained earnings were non stationarity at level since p-value>0.05. This means that the results obtained are now not spurious (Gujarati, 2009) and so panel regression models could be generated.

**4.4.2 Normality Test**
The normality assumption ($u_t \sim N(0, \sigma^2)$) was required in order to conduct single or joint hypothesis tests about the model parameters (Brooks, 2008). Table 5 shows the normality results using for Skewness and Kurtosis test for the financial firms. Bera and Jarque (1981) tests of normality were performed. If the p-value is less than 0.05, the null of normality at the 5% level is rejected. If the data is not normally distributed a non-parametric test will be most appropriate. The study tested the null hypothesis that, the disturbances are not normally distributed.
$H_0$: The data are not normally distributed
$H_1$: The data are not normally distributed

**Table 5: Normality Test**

| Variable | Observation | Skewness | Kurtosis | p-value |
|---|---|---|---|---|
| Earnings Per Share | 360 | 1.0670 | 0.7324 | .166 |
| Market Capitalization | 360 | 3.3921 | 0.9205 | .453 |
| Short Term Debt | 360 | 2.0211 | 0.6413 | .825 |

Table 5 shows the normality results using for skewness and Kurtosis test. The P-values were higher than the critical 0.05 and thus we conclude that the data is normally distributed.

### 4.4.3 Multicollinearity Test

According to Daoud (2017), multicollinearity refers to the presence of correlations between the predictor variables. In severe cases of perfect correlations between predictor variables, multicollinearity can imply that a unique least squares solution to a regression analysis cannot be computed (Alin, 2010). Multicollinearity inflates the standard errors and confidence intervals leading to unstable estimates of the coefficients for individual predictors (Daoud, 2017). Multicollinearity was assessed in this study using the variance inflation factors (VIF). According to Alin (2010), VIF values in excess of 10 is an indication of the presence of Multicollinearity. The results in Table 6 indicated absence of multicollinearity since the VIF of all the variables were less than 10.

**Table 6: Multicollinearity Test**

|  | Growth in Earnings Per Share | | Growth in Market Capitalization | |
|---|---|---|---|---|
| Variable | 1/VIF | VIF | 1/VIF | VIF |
| Short Term Debt | 1.03 | 0.967385 | 1.25 | 0.800058 |

The results in Table 6 indicated absence of multicollinearity since the VIF of all the variables were less than 10. When multicollinearity was tested with growth in earning per share as a measure of financial growth, the VIF values for short-term debt were less than 10. Likewise, when multicollinearity was tested with growth in market capitalization as a measure of financial growth, the VIF values for short-term debt were less than 10 indicating absence of multicollinearity.

### 4.5 Panel Regression Analysis Results
### 4.5.1 Effect of Short-Term Debt on Financial Growth

Random effect model was estimated between short term debt and measures of financial growth (growth in earnings per share and growth in market capitalization). Panel regression was conducted to determine whether there was a significant relationship between short term debt and growth in earning per share. Table 7 presents the regression model on short term debt with growth in earning per share as a measure of financial growth.

**Table 7: Effect of Short-Term Debt on Growth in Earning Per Share (EPS)**

| Growth in EPS | Coef. | Std. Err. | T | P>|t| | [95% Conf. | Interval] |
|---|---|---|---|---|---|---|
| Short Term Debt | 0.038201 | 0.009365 | 4.08 | 0.000 | 0.019847 | 0.056555 |
| _cons | 2.731963 | 1.198171 | 2.28 | 0.023 | 0.383591 | 5.080334 |
| R-squared: | 0.4599 | | | | | |
| Wald chi2(1) | 16.64 | | | | | |
| Prob | 0.0000 | | | | | |

The fitted model from the result is
Growth in EPS = 2.731963 + 0.038201STD
Where: EPS = Earnings Per Share
STD = Short Term Debt

As presented in the Table 7, the coefficient of determination R-Square is 0.4599. The model indicates that, short term debt explains 45.99% of variation in growth in earnings per share as a measure of financial growth. This means 45.99% of variations in the growth in earning per share is influenced by short term debt. The findings further confirm that, the relationship between short term debt and growth in earnings per share as a measure of financial growth is positive and significant with a coefficient of (β=0.038201, p=0.000). This implies that, there exist a positive and significant relationship between short term debt and growth in earnings per share as a measure of financial growth since the coefficient value was positive and the p-value 0.000<0.05. This means that a unitary increase in short term debt leads to an increase in growth in earnings per share by 0.038201 units holding other elements of financial structure constant. The results agree with García and Martínez (2008) analyzed the Spanish Small and Medium Enterprises (SMEs) cash holdings and found that, firms with a higher amount of short-term debt will hold higher levels of cash, because it might lower the risks of non-renewal of the short-term debt and that short-term debt is positively related to firm's profitability measured using earnings per share. The results, however, do not agree with Tifow and Sayilir (2015) that short term debt has a significant negative relationship with earnings per share. The results are also inconsistent with Salim and Yadav (2012) investigated the relationship between capital structure and firm performance and established that short term debt has a negative relationship with earnings per share. Table 5 presents the regression model on short term debt versus growth of market capitalization as a measure of financial growth.

**Table 5: Effect of Short-Term Debt on Growth in Market Capitalization**

| Growth of Market Capitalization | Coef. | Std. Err. | T | P>|t| | [95% Conf. | Interval] |
|---|---|---|---|---|---|---|
| Short Term Debt | 0.042264 | 0.009848 | 4.29 | 0.000 | 0.022962 | 0.061566 |
| _cons | 6.508236 | 1.260041 | 5.17 | 0.000 | 4.038602 | 8.977871 |
| R-squared: | 0.2560 | | | | | |
| Wald chi2(1) | 18.42 | | | | | |
| Prob | 0.0000 | | | | | |

The fitted model from the result is

Growth in Market Capitalization = 6.508236 + 0.042264STD

Where: STD = Short Term Debt

As presented in the Table 5, the coefficient of determination R-Square is 0.2560. The model indicates that, short term debt explains 25.6% of variations in the growth in market capitalization as a measure of financial growth. This means 25.6% of variation in the growth in market capitalization is influenced by short term debt. The findings further confirm that, the regression model for short term debt and growth in market capitalization as a measure of financial growth is positive and significant with a coefficient of ($\beta$ =0.042264, p=0.000). This implies that, there exist a positive and significant relationship between short term debt and growth in market capitalization as a measure of financial growth since the coefficient value was positive and the p-value 0.000<0.05. This means that a unitary increase in short term debt leads to growth in market capitalization by 0.042264 units holding other elements of financial structure constant. Short-term debt is less expensive than long-term debt but is riskier because they need to be renewed periodically. A firm may find itself in a crisis if they are unable to renew their debt usually because of some negative news, real or otherwise. Most failures of large corporations are precipitated by the unavailability of short-term funding. The results are consistent with Garcia-Terul and Martinez -Solano (2008) that short-term debt is positively correlated with firm's growth opportunities. According to Fosberg (2013) a significant decrease in asset sales contributes to the need for more short-term debt financing.

**4.6 Hypothesis Testing**
The hypothesis was tested using p-value method. The acceptance/rejection criterion was that, if the p-value is greater than the significance level of 0.05, we fail to reject the Ho but if it's less than 0.05 level of significance, the Ho is rejected. The null hypothesis was that, there is no significant effect of short-term debt on financial growth of Non-financial firms listed at Nairobi Securities Exchange. The results in Table 4 shows that, short term debt and growth in earning per share are positively and significantly related with p-value =0.000<0.05. Further, the results in Table 5 shows that, short term debt and growth in market capitalization as a measure of financial growth is positive and significantly related with p-value =0.000<0.05. The null hypothesis was therefore rejected and concluded that, there is a significant effect of short-term debt on financial growth of Non-financial firms listed at Nairobi Securities Exchange.

**5.0 CONCLUSION**
Based on the findings, the study concluded that, short term debt has a positive and significant relationship with financial growth measured using growth in earning per share. Earnings per share is considered as an important accounting indicator of risk, entity performance and corporate success. It is used to forecast potential growth in future share prices, because changes in earning per share are often reflected in share price behavior. Earnings per share is a useful investment decision tool for investors, because it indicates future prospects and growth. Short-term assets should be financed with short-term liabilities and long-term assets should be financed with long-term liabilities. Short-term term is primarily concerned with the analysis of decisions that affect current assets and current liabilities.

It was also concluded that, short term debt has a positive and significant relationship with financial growth as measured by growth in market capitalization. Growth in market capitalization is important in projecting the size of the firm because it shows the firm's value. Market capitalization is a measure of the value of companies and stock markets which is an on-going market valuation of a public firm whose shares are publicly traded on a stock exchange computed by multiplying the number of outstanding shares held by the shareholders with the current per share market price at a given time. Short-term debt could be used as permanent source of financing if the debt is continually refinanced as it matures. One reason to use short-term debt as a permanent source of financing is to take advantage of an upward sloping yield curve to reduce the firm's interest expense.

**6.0 RECOMMENDATIONS**
The study revealed that, short term debt influences financial growth. The study recommends that, the management of non-financial firms listed at Nairobi Securities Exchange to employ financing means that can improve the earnings per share, market capitalization and enhance the value of the firm for the benefit of its stakeholders.

**REFERENCES**

Abdeldayem, M. M. (2015). Examining the Relationship between Agency Costs and Stock Mispricing: Evidence from the Bahrain Stock Exchange. *International Journal of Economics, Commerce and Management*, *3*(4), 1-35.

Aburub, N. (2012). Capital Structure and Firm Performance: Evidence from Palestine Stock Exchange. *Journal of Money, Investment and Banking*, 23(1), 109-117.

Akbarpour, M., & Aghabeygzadeh, S. (2011). Reviewing Relationship between financial structure and firm's performance in firms traded on the Tehran Stock Exchange. International Journal of Business Administration, 2(4), 175.

Alin, A. (2010). Multicollinearity. Wiley Interdisciplinary Reviews: Computational Statistics, 2(3), 370-374.

Baum, C. F., Schafer, D., & Talavera, O. (2006). *The effects of short-term liabilities on profitability: a comparison of German and US firms* (Vol. 636). Boston College Working Papers in Economics.

Bera, A. K., & Jarque, C. M. (1981). Efficient tests for normality, homoscedasticity and serial independence of regression residuals: Monte Carlo evidence. Economics letters, 7(4), 313-318.

Bryman, A., & Bell, E. (2011). Ethics in business research. *Business Research Methods*, *7*(5), 23-56.

Buferna, F. M., Bangassa, K., & Hodgkinson, L. (2005). *Determinants of capital structure: evidence from Libya* (Vol. 8). University of Liverpool.

Buvanendra, S., Sridharan, P., & Thiyagarajan, S. (2017). Firm characteristics, corporate governance and capital structure adjustments: A comparative study of listed firms in Sri Lanka and India. *IIMB management review*, *29*(4), 245-258.

Chen, J. J. (2014). Determinants of capital structure of Chinese-listed companies. *Journal of Business research*, *57*(12), 1341-1351.

 Cost Theory Perspective. *American International Journal of Social Science, 3*(1), *139-140.*

Daoud, J. I. (2017, December). Multicollinearity and regression analysis. In *Journal of Physics: Conference Series (Vol. 949, No. 1, p. 012009). IOP Publishing.*

Dare, F.D., & Sola, O. (2010). Capital Structure and Corporate Performance in Nigeria Petroleum Industry: Panel Data Analysis*. Journal of Mathematics and Statistics 6*(2), 168-173.

Dawar, V. (2014). Agency theory, capital structure and firm performance: some Indian evidence. *Managerial Finance*, *40*(12), 1190-1206.

Diamond, D. W., & He, Z. (2014). A theory of debt maturity: the long and short of debt overhang. *The Journal of Finance*, *69*(2), 719-762.

Diaz Hermelo, F. (2007). The determinants of firm's growth: an empirical examination.

Fosberg, R. H. (2013). Short-term debt financing during the financial crisis. *International journal of business and social science*, *4*(8), 45-57.

Gambacorta, L., Yang, J., & Tsatsaronis, K. (2014). Financial structure and growth. BIS *Quarterly Review*, March 2014. Available at SSRN: https://ssrn.com/abstract=2457106.

García, P. J., & Martínez, P. (2008). On the determinants of SME cash holdings: Evidence from Spain. Journal of Business Finance & Accounting, 35(1-2), 127-149.

García-Teruel, P. J., & Martínez-Solano, P. (2007). Short-term debt in Spanish SMEs. International Small Business Journal, 25(6), 579-602.

Ghazouani, T. (2013). The capital structure through the trade-off theory: evidence from Tunisian firm. *International Journal of Economics and Financial Issues*, *3*(3), 625-636.

Githaiga, P. N., & Kabiru, C. G. (2015). Debt financing and financial performance of small and medium size enterprises: evidence from Kenya. *Journal of Economics, Finance and Accounting*, *2*(3), 473-481.

Guin, L. (2011). Matching Principle. Murray State University, Tutorial.

Gujarati, D. N. (2009). Basic econometrics. Tata McGraw-Hill Education.

Habib, H., Khan, F., & Wazir, M. (2016). Impact of Debt on Profitability of Firms: Evidence from Non-Financial Sector of Pakistan. *City University Research Journal*, Volume 06, Number 01, January 2016. Available at SSRN: https://ssrn.com/abstract=2714461

Heinkel, R. (1982). A theory of capital structure relevance under imperfect information. *The journal of finance*, *37*(5), 1141-1150.

Ishaya, L. C., & Abduljeleel, B. O. (2014). Capital Structure and Profitability of Nigerian Quoted: The Agency

Jensen, M. C., & Meckling, W. H. (1976). Theory of the firm: managerial behaviour, agency costs and ownership structure. *Journal of Financial Economics*, 3, 305 - 360.

Kahl, M., Shivdasani, A., & Wang, Y. (2015). Short-Term Debt as Bridge Financing: Evidence from the Commercial Paper Market. *The Journal of Finance*, *70*(1), 211-255.

Koila, E., Kiru, K., & Koima, J., K. (2018). Relationship between Share Market Capitalization and Performance of Listed Firms at Nairobi Securities Exchange Limited, Kenya Relationship between Share Market, *Scholars

*Journal of Economics, Business and Management, 5(3), 1-23*

Krishnamurthy, A., & Vissing-Jorgensen, A. (2013). Short-term debt and financial crises: What we can learn from US Treasury supply. *Unpublished, Northwestern University, May 2012.*

Kumah, S. P. (2013). Corporate Capital Structure Determinants of Listed Firms in West African Monetary Zone– A Review of Related Literature. *Research Journal of Finance and Accounting*, *4*(19), 19-29.

Leland, H. E. (1998). Agency costs, risk management, and capital structure. *The Journal of Finance*, *53*(4), 1213-1243.

Lixin, X., & Lin, C. (2008). The relationship between debt financing and market value of company: empirical study of listed real estate company of China. In *7th international conference on innovation & management, China, August 2008, 2043-2047*.

Ma'aji, M. M., Abdullah, N. A. H., & Khaw, K. L. H. (2018). Predicting financial distress among SMEs in Malaysia. *European Scientific Journal*, *14*(7), 91-102.

Magoro, K., & Abeywardhana, D. (2017). Debt capital and financial performance: A study of South African companies. *International Journal of Scientific Research and Innovative Technology*, *4*(4), 71-84.

McMillan, J.H. & Schumacher, S. (2006). Research in education: Evidence based inquiry, *(6th ed.) Boston, United States of America: Pearson Education.*

Menike M. G. P. & U. S. Prabath (2014). The Impact of Accounting Variables on Stock Price: Evidence from the Colombo Stock Exchange, Sri Lanka: *International Journal of Business and Management*; 9 (5), 210–221.

Mlonzi, V.F., Kruger, J. & Ntoesane, M.G. (2011). Share price reaction to earnings announcement on the JSE-AltX: A test for mark efficiency. *Southern African Business review*, 15(3), 142-166.

Muchiri, M. J., Muturi, W. M., & Ngumi, P. M. (2016). Relationship between Financial Structure and Financial Performance of Firms Listed at East Africa Securities Exchanges. *Journal of Emerging Issues in Economics, Finance and Banking,* 7(5), 1734-1755.

Muchugia, L. M. (2013). The effect of debt financing on firm profitability of commercial banks in Kenya. Unpublished MBA project, University of Nairobi.

Mwangi, L. W., Muathe, S. M. A., & Kosimbei, G. K. (2014). Relationship between capital structure and performance of non-financial companies listed in the Nairobi Securities Exchange, Kenya.

NSE report (2016). NSE Annual Report 2016 - Nairobi Securities Exchange. Available at www.nse.co.ke.

NSE report (2017). NSE Annual Report 2017 - Nairobi Securities Exchange. Available at www.nse.co.ke.

O'Neill, P., Sohal, A., & Teng, C. W. (2016). Quality management approaches and their impact on firms′ financial performance - An Australian study. *International Journal of Production Economics*, (17) 1, 381-393.

Oma, M., D & Memba, F., S. (2018). The Effect of Share Capital Finance on Profitability of Petroleum Marketing Firms in Kenya, *International Journal of Economics, Commerce and Management*,6(1),410-422

Onaolapo, A.A., & Kajola, S.O. (2010). Capital Structure and Firm Performance: Evidence from Nigeria. *European Journal of Economics, Finance and Administrative Sciences,* 25 (7), 70-82.

Ongore, V. O., & Kusa, G. B. (2013). Determinants of financial performance of commercial banks in Kenya. *International Journal of Economics and Financial Issues*, *3*(1), 237-249.

Opungu, J. A. (2016). *The Effect of Capital Structure on Profitability of Non-Financial Firms Listed at Nairobi Security Exchange* (Doctoral dissertation, KCA University).

Peavler, R. (2014). Profitability ratio analysis. Retrieved June 6, 2016.

Peirson, G., Brown, R., Easton, S., & Howard, P. (2014). *Business finance*. McGraw-Hill Education Australia.

Penrose, E. T. (1959). The theory of the growth of the firm. *New York: Sharpe*.

Pervan, M. & Visic, J. (2012). Influence of firm size on its business success: *Croatian Operational Research Review (CRORR),* 3(1), 213-216.

Rajan, R. G., & Zingales, L. (1998). Power in a Theory of the Firm. The Quarterly Journal of Economics, 113(2), 387-432.

Rashid, A. (2015). Revisiting agency theory: Evidence of board independence and agency cost from Bangladesh. *Journal of business ethics*, *130*(1), 181-198.

Robbetze, N., de Villiers, R., & Harmse, L. (2017). The Effect of Earnings Per Share Categories on Share Price Behavior: Some South African Evidence. *Journal of Applied Business Research (JABR)*, *33*(1), 141-152.

Salim, M., & Yadav, R. (2012). Capital structure and firm performance: Evidence from Malaysian listed companies. *Procedia-Social and Behavioral Sciences, 65, 156-166.*

San, O.T., & Heng, T.B. (2011). Capital Structure and Corporate Performance of Malaysian Construction Sector. *International Journal of Humanities and Social Science,* 1(2), 28-36.

Shubita, M. F., & Alsawalhah, J.F. (2012). The Relationship between Capital Structure and Profitability. *International Journal of Business and Social Science, 3*(16), 33-44.

Smart, S.B. & Graham, J.R. (2012*). Introduction to financial management*. (3rd ed.). Belmont, CA: Cengage Learning.

Subedi, D. (2016). Explanatory sequential mixed method design as the third research community of knowledge claim. *American Journal of Educational Research*, *4*(7), 570-577.
Tailab, M. (2014). The effect of capital structure on profitability of energy American firms. *International Journal of Business and Management Invention, 3(12)*.
Tifow, A. A., & Sayilir, O. (2015). Capital structure and firm performance: An analysis of manufacturing firms in Turkey. *Eurasian Journal of Business and Management, 3(4), 13-22.*
Yazdanfar, D., & Öhman, P. (2015). Debt financing and firm performance: an empirical study based on Swedish data. *The Journal of Risk Finance*, *16*(1), 102-118.